# Superconducting properties and magneto-optical imaging of $Ba_{0.6}K_{0.4}Fe_2As_2$ PIT wires with Ag addition

Qing-Ping Ding[1,2], Tritat Prombood[1], Yuji Tsuchiya[1], Yasuyuki Nakajima[1,2] and Tsuyoshi Tamegai[1,2]

[1]Department of Applied Physics, The University of Tokyo, 7-3-1 Hongo, Bunkyo-ku, Tokyo 113-8656, Japan
[2]JST, Transformative Research-Project on Iron Pnictides (TRIP), 7-3-1 Hongo, Bunkyo-ku, Tokyo 113-8656, Japan

**Abstract.** We have fabricated $(Ba,K)Fe_2As_2$ superconducting wires through *ex-situ* powder-in-tube method. Silver was used as a chemical addition to improve the performance of these superconducting wires. The transport critical current densities ($J_c$) have reached $1.3 \times 10^4$ A/cm and $1.0 \times 10^4$ A/cm$^2$ at 4.2 K under self field in the wires with and without Ag addition. We used magneto-optical (MO) imaging technique to investigate the properties of grain boundaries in the $(Ba,K)Fe_2As_2$ superconducting wire with Ag addition. MO images show the weak links in the Fe-based superconducting wires for the first time. An intragranular $J_c$ of $6.0 \times 10^4$ A/cm$^2$ at 20 K is obtained from MO image, which is consistent with the estimation from *M-H* measurement.

Applications of superconducting wires include high power density power cables, transformers, and superconducting magnetic energy-storage devices for power networks where low reactance and instantaneous redistribution of power is important [1]. To allow the flow of high superconducting currents the vortices need to be pinned. This is quite feasible at low temperature, but the thermal fluctuations make it difficult at high temperatures, which will be worsened by the anisotropy of the layered compounds. Iron-based superconductor $(Ba,K)Fe_2As_2$ has a low anisotropy [2, 3]. The upper critical field, $H_{c2}(T)$, for $(Ba,K)Fe_2As_2$ exceed both $MgB_2$ and $Nb_3Sn$ at low temperatures. Now the target of $MgB_2$ wire technology is to develop magnetic resonance imaging magnets that can operate at 20 K instead of 4.2 K. With $H_{c2}$ of nearly 40 T at 20 K, $(Ba,K)Fe_2As_2$ attracts great technological interest [4]. Superconducting wires of iron-based superconductors have been fabricated [5-7] soon after their discoveries [8, 9]. The performances of the iron-based superconducting wires have been much improved by addition of Ag, Pb and Sn [6, 10, 11]. In this paper, we investigated the effect of Ag addition on the performance of $(Ba,K)Fe_2As_2$ wires. We used magneto-optical (MO) imaging technique (12, 13) to investigate the properties of grain boundaries and the intragranular critical current densities ($J_c^{intra}$) in the wire. Usually, weak links between superconducting grains have been considered to be the main reasons for the low transport $J_c$ and its strong field dependence in iron-based superconducting wires. Our MO images demonstrate the weak link behavior in these wires for the first time.

$Ba_{0.6}K_{0.4}Fe_2As_2$ polycrystalline samples were prepared by solid-state reaction method. We used Ba pieces (Kojundo Chemical Laboratory, 99%), K ingots (same as above, 99.5%), and FeAs powder as raw materials. FeAs was prepared by placing stoichiometric amounts of As pieces (Fukuzawa Electric, 99.99999%) and Fe powder (Kojundo Chemical Laboratory, 99%) in an evacuated quartz tube and reacting them at 1065 °C for 10 h after heating at 700 °C for 6 h. A mixture with a ratio of Ba: K: FeAs = 0.6: 0.44: 2 was placed in an alumina crucible. The whole assembly was sealed in a quartz tube, and slowly ramped up to 1100 °C in 20 h followed by cooling down to room temperature naturally. 10% K was added in order to compensate the reaction with quartz tube at high temperatures [10, 14].

Powder-in-tube (PIT) method was adopted to fabricate $Ba_{0.6}K_{0.4}Fe_2As_2$ superconducting wires. As-prepared $Ba_{0.6}K_{0.4}Fe_2As_2$ polycrystalline sample was ground into fine powder with an agate mortar and pestle in nitrogen-filled glove box. The powder was divided into two parts. One part was filled into Ag tube with OD 4.5 mm and ID 3 mm (The Nilaco Corpoartion, 99%) directly. The other part was mixed with 15% in weight of Ag powder (Kojundo Chemical Laboratory, 99.9%) and then put into Ag tube with the same dimension. The assembly was cold drawn into a square wire with diagonal dimension of about 0.6 mm. The as-drawn wire was cut into shorter wires with length ~ 40 mm, and then sealed in an evacuated quartz tube. The sealed wires were put into a muffle furnace and heated up to temperatures between 600°C to 900°C with a ramping rate of 100 °C/h, and kept at this temperature for 12 h to 36 h. Then the furnace was switched off and cooled to room temperature naturally.

The phase identification of the sample was carried out by means of powder X-ray diffraction (M18XHF, MAC Science) with Cu-$K\alpha$ radiation generated at 40kV and 200 mA. Bulk magnetization is measured by a superconducting quantum interference device (SQUID) magnetometer (MPMS-5XL, Quantum Design). Resistivity and current-voltage (*I-V*) measurements were performed by the four-probe method with silver paste for contacts. Resistivity measurements were performed within the sample chamber of a SQUID magnetometer. *I-V* measurements were performed in a bath-type cryostat (Spectromag, Oxford Instruments). For MO imaging, the wire was cut using a wire saw, and the surface was polished with a lapping film. A Bi-substituted iron-garnet indicator film is placed in direct contact with the sample, and the whole assembly is attached to the cold finger of a He-flow cryostat (Microstat-HR, Oxford Instruments). MO images are acquired by using a cooled CCD camera with 12-bit resolution (ORCA-ER, Hamamatsu). To enhance the visibility of the local magnetic induction and eliminate the signals from the impurity phases and scratches in the garnet film, a differential imaging technique is employed [15, 16].

Figure 1(a) shows the X-ray diffraction pattern of as-prepared $Ba_{0.6}K_{0.4}Fe_2As_2$ polycrystalline sample. All the peaks can be well indexed using a space group of I4/mmm except for small peaks from impurity phases marked by asterisks. The compound crystallizes in a tetragonal structure and the majority of the impurity phases are identified as remaining FeAs. Temperature dependences of zero-field-cooled (ZFC) and field-cooled (FC) magnetization of the $Ba_{0.6}K_{0.4}Fe_2As_2$ polycrystalline sample are shown in the inset of Fig. 1(a). The sample shows an onset of diamagnetism at $T_c$ ~ 37 K. The transition temperature is similar to that of an optimally-doped $(Ba,K)Fe_2As_2$ [6]. Figure 1(b) shows the X-ray pattern of the core of the wire with Ag addition after sintered at 700°C for 24h. All the measurements described below were performed on this wire. Compared with the polycrystalline sample before filling into the

Ag tubes, several new peaks which were assigned to the added Ag were detected. Shown in the inset of Fig. 1(b) are the temperature dependences of magnetization of the sintered wire with Ag addition. The $T_c$ has decreased to 34 K after sintering.

Fig. 2(a) shows the transverse cross section of the sintered wire with Ag addition. We did not observe a reaction between $Ba_{0.6}K_{0.4}Fe_2As_2$ core and Ag sheath, thus demonstrating the stability of Ag as a good sheath material for $(Ba,K)Fe_2As_2$. A higher magnification optical image of the core is shown in Fig. 2(b). The very bright spots correspond to the added Ag powder, and the dark regions are the voids which were produced during filling and drawing process.

We investigated the transport $J_c$ of those superconducting wires. The inset of Fig. 3(a) shows the $E-J$ characteristics of the wire with Ag addition under different fields up to 90 kOe at 4.2 K. Here we adopt $E = 1$ μV/cm for the $E-J$ curve as a criterion to define transport $J_c$. Transport $J_c$ as a function of field at 4.2 K for both wires with and without Ag addition are shown in Fig. 3(a). Compared with the wire without Ag addition, the transport $J_c$ was enhanced by Ag addition. The transport $J_c$ is $1.3 \times 10^4$ A/cm$^2$, and $1.0 \times 10^4$ A/cm$^2$ at 4.2 K under zero field for the wire with and without Ag addition, respectively. Not only the self-field $J_c$, but also its strong field dependence has been improved remarkably through Ag addition, especially at low fields (below 20 kOe). We tried to sinter the wires at several different temperatures with different periods, and found 700°C for 24h is the optimal condition, which provides the highest transport $J_c$ for both wires with and without Ag addition. Below 600 °C the transport $J_c$ is just several hundred A/cm$^2$. Above 900°C $Ba_{0.6}K_{0.4}Fe_2As_2$ core will react with Ag sheath even it has not reached the melting point of Ag (961 °C). The self-field transport $J_c$ of pure $Ba_{0.6}K_{0.4}Fe_2As_2$ superconducting wire is the highest among all the reported pure iron-based superconducting wires so far [6, 10, 11]. This indicates that $Ba_{0.6}K_{0.4}Fe_2As_2$ may be the best candidate for iron-based superconducting wires. By further optimization like texturing, the performance of $Ba_{0.6}K_{0.4}Fe_2As_2$ superconducting wire should be enhanced much more [11]. Temperature dependence of resistance under magnetic fields from 0 to 50 kOe is shown in Fig. 3 (b). Inset of Fig. 3 (b) shows the variation of the upper critical field $H_{c2}$ with temperature for the wire with Ag addition. The values of $H_{c2}$ are defined as the field at the midpoint of the resistive transition. The slope of $H_{c2}$ at $T_c$ is -27.3 kOe/K. The value of $H_{c2}$ at $T = 0$ K estimated using the Werthamer–Helfand–Hohenberg formula [17], $H_{c2}(0)=-0.69T_c|dH_{c2}/dT|_{T=T_c}$, is 595 kOe. This value is a little smaller than that of the single crystal [18].

In iron-based superconductors, the vortices at grain boundaries are generally pinned more weakly than the vortices in the grains, and the grain boundaries become barriers to current flow. From the transport $J_c$, our $Ba_{0.6}K_{0.4}Fe_2As_2$ wire with Ag addition is one of the best iron-based superconducting wires [10, 11]. So we want to investigate the quality of the grain boundaries in this wire. Figures 4(a)-4(c) depict MO images of the transverse cross section of the wire with Ag addition in the remanent state after applying a 500 Oe field for 0.25 seconds which was subsequently reduced down to zero at different temperatures. The thickness of this sample cut perpendicular to the wire axis for MO imaging is roughly 200μm. The bright regions correspond to the trapped flux in the sample. These images are similar to the MO images of 1111 and 11 polycrystals and tapes [19-24]. At all temperatures, bright intensities are restricted in small regions, implying that the intergranular current density is much smaller compared with the intragranular current density. The intragranular current density decreases gradually as the temperature is increased towards $T_c$. Fig. 4(d) shows the magnetic induction profiles along the dashed line in Fig. 4(a). From the magnetic induction profile we calculated the $J_c^{intra}$. In this calculation, we roughly estimate the intragranular critical current density by $J_c^{intra} \sim dB/dx$. The estimated $J_c^{intra}$ is $\sim 6.0 \times 10^4$ A/cm$^2$ at 20 K for relatively large grains.

Magnetic hysteresis curves of the wire are shown in Fig. 5(a). By using the Bean model with an assumption of field–independent $J_c$, intragranular critical current density $J_c^{intra}$ for polycrystalline samples can be evaluated from the magnetization hysteresis loops [25]. According to the Bean model, $J_c^{intra}$[A/cm$^2$] is given by $J_c^{intra} = 30 \Delta M/d$, with an assumption that intergranular critical current is zero, where $\Delta M$[emu/cc] is $M_{down} - M_{up}$, $M_{up}$ and $M_{down}$ are the magnetization when sweeping field up and down, respectively, $d$[cm] is the effective diameter of the grains in the polycrystalline sample. If we assume that grains with different sizes occupy the same volume, the effective diameter $d$ is $\sim 1/2\ d_{max}$, here $d_{max}$ is the largest grain size in the assumed distribution. From the optical image in Fig. 2(b), the largest grain size is $\sim 20$ μm, which means $d$ is $\sim 10$ μm. Temperature dependence of $J_c^{intra}$ of the wire with Ag addition estimated from $M-H$ and MO measurements are shown in Fig. 5(b). $J_c^{intra}$ estimated from $M-H$ curve is consistent with the value estimated from MO measurements.

Figures 6(a) to 6(d) reveal the penetration of magnetic flux at 5 K in the wire with Ag addition. Magnetic flux penetrates intergranular regions even when the field is small (Fig. 6(a)). When the field is increased further (Figs. 6(b)-6(d)), the shapes of the superconducting grains are more visible. These images demonstrate the effect of weak links between superconducting grains even at low fields.

In summary, we performed X-ray diffraction, magnetization, resistivity, transport critical current density and magneto-optical measurements on $Ba_{0.6}K_{0.4}Fe_2As_2$ superconducting wires with Ag addition prepared through PIT method. Transport $J_c$ of $1.3\times10^4$ A/cm and $1.0\times10^4$ A/cm$^2$ at 4.2 K under self field have been obtained in the wires with and without Ag addition. MO images show the weak links in the Fe-based superconducting wires for the first time.

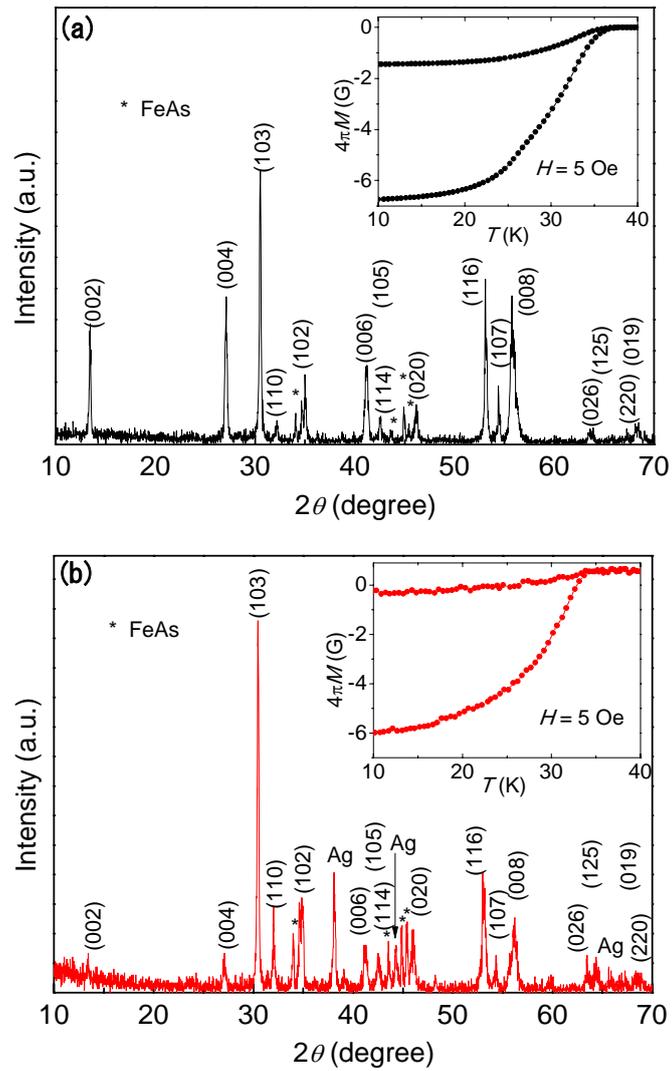

**Figure 1.** (a) Powder X-ray diffraction pattern of $Ba_{0.6}K_{0.4}Fe_2As_2$ polycrystalline sample. Inset: Temperature dependences of magnetization of the polycrystalline sample. (b) Powder X-ray diffraction pattern of $Ba_{0.6}K_{0.4}Fe_2As_2$ PIT superconducting wire with Ag addition. Inset: Temperature dependences of magnetization of the wire.

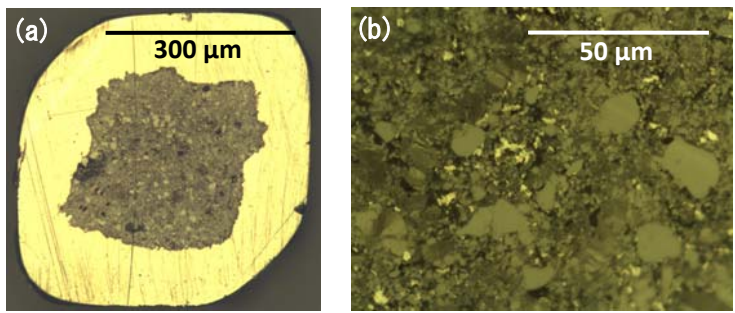

**Figure 2.** (a) Optical images of the transverse cross section for $Ba_{0.6}K_{0.4}Fe_2As_2$ PIT superconducting wires. (b) Higher magnification optical image of the core part in (a).

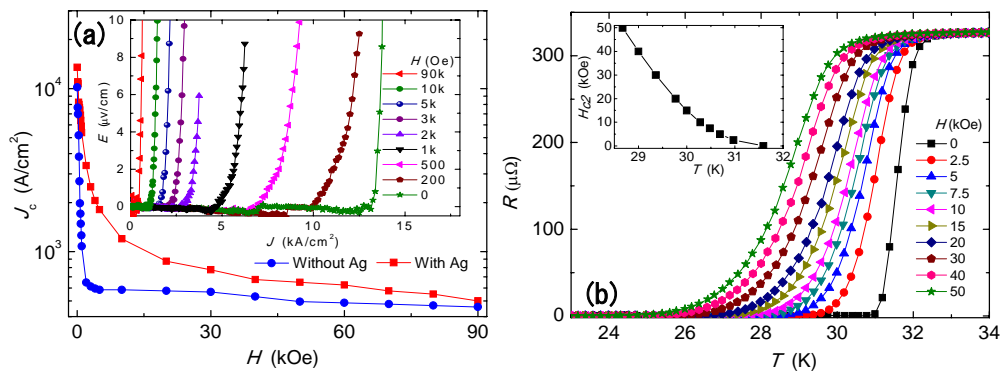

**Figure 3**. (a) Magnetic field dependences of the transport $J_c$'s in $Ba_{0.6}K_{0.4}Fe_2As_2$ PIT superconducting wires with and without Ag addition. Inset shows $E$–$J$ characteristics in the wire with Ag addition under different fields. (b) Field dependence of the resistivity of the wire with Ag addition around $T_c$. Inset shows the upper critical field $H_{c2}$ versus temperature.

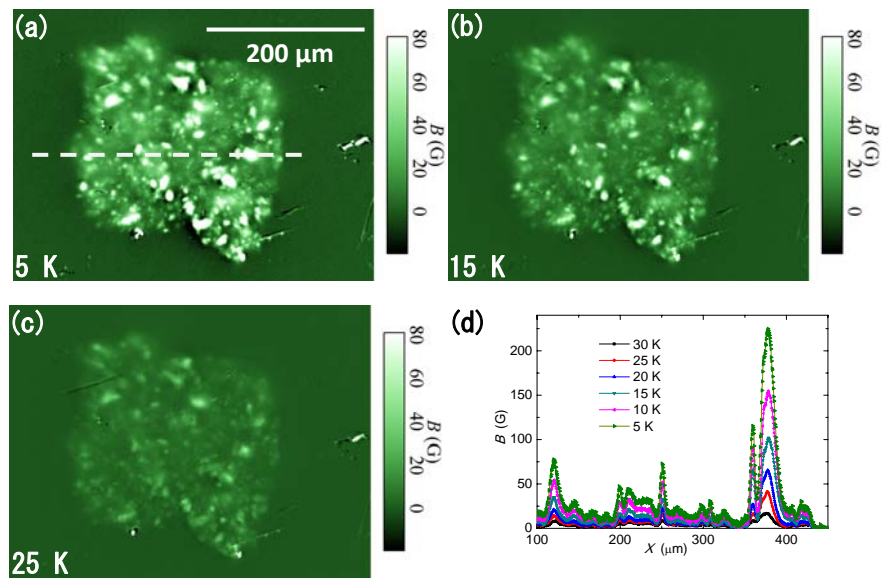

**Figure 4.** Differential MO images in the remanent state of the $Ba_{0.6}K_{0.4}Fe_2As_2$ PIT superconducting wires at (a) 5 K, (b) 15 K, and (c) 25 K after cycling the field up to 500 Oe for 0.25 seconds. (d) The local magnetic induction profiles at different temperatures taken along the dashed lines in (a).

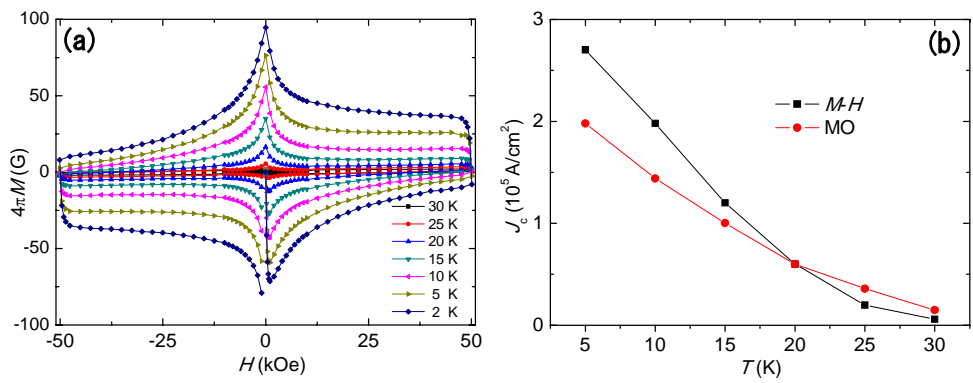

**Figure 5.** (a) Magnetic field dependence of magnetization of the wire with Ag addition. (b) Temperature dependence of intragranular $J_c$ of the wire with Ag addition estimated from *M-H* and MO measurements.

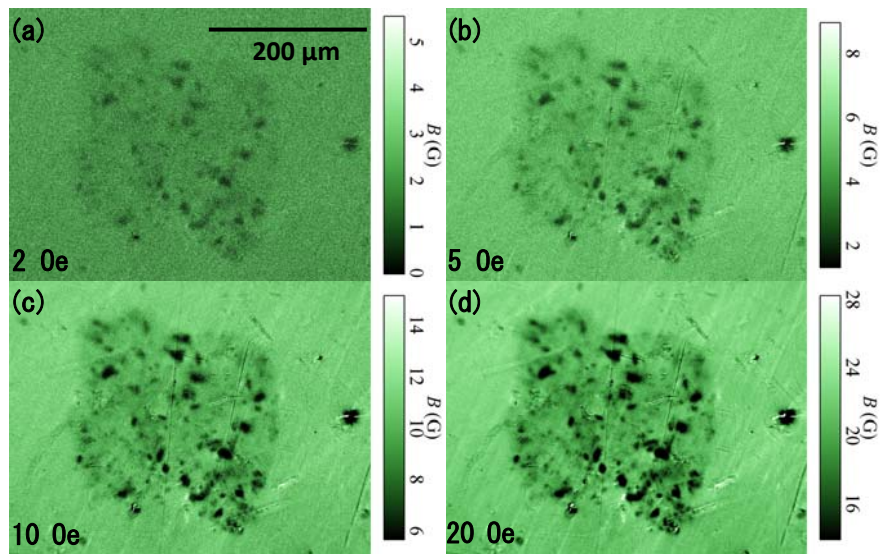

**Figure 6.** MO images of flux penetration into $Ba_{0.6}K_{0.4}Fe_2As_2$ PIT superconducting wire with Ag addition at (a) 2 Oe, (b) 5Oe, (c) 10 Oe, and (d) 20 Oe after zero-field cooling down to 5 K.